\documentclass[letterpaper]{article} 
\usepackage{aaai2026}  
\usepackage{times}  
\usepackage{helvet}  
\usepackage{courier}  
\usepackage[hyphens]{url}  
\usepackage{graphicx} 
\usepackage{natbib}  
\usepackage{caption} 
\usepackage{algorithm}
\usepackage{algorithmic}

\usepackage[utf8]{inputenc} 
\usepackage[T1]{fontenc}    
\usepackage{url}            
\usepackage{booktabs}       
\usepackage{amsfonts}       
\usepackage{nicefrac}       
\usepackage{microtype}      
\usepackage{xcolor}         
\usepackage{enumitem}
\usepackage{graphicx}
\usepackage{multirow}

\usepackage{tcolorbox} 
\usepackage{listings} 

\usepackage[frozencache,cachedir=minted-cache]{minted}
\usepackage{xcolor} 
\usepackage{caption}
\usepackage{enumitem}
\tcbuselibrary{listings, breakable}
\definecolor{lightorange}{RGB}{253,188,180}
\definecolor{lightblue}{RGB}{180,208,253}
\definecolor{problemcolor}{RGB}{153,0,153}
\definecolor{modelcolor}{RGB}{0, 0, 255}

\definecolor{pythonblue}{RGB}{84, 184, 255}
\definecolor{soappink}{RGB}{255, 163, 203}

\newtcolorbox{DS1000Box}[1][]{
    colback=soappink!5!white,
    colframe=soappink!75!black,
    title=\textbf{Task Description},
    top=2mm,
    bottom=2mm,
    left=2mm,
    right=2mm,
    arc=2mm, 
    boxsep=1mm,
    boxrule=0mm,
    breakable,
    #1
}

\newtcolorbox{DSCodeBenchBox}[1][]{
    colback=pythonblue!5!white,
    colframe=pythonblue!75!black,
    title=\textbf{Model},
    top=2mm,
    bottom=2mm,
    left=2mm,
    right=2mm,
    arc=2mm,
    boxsep=1mm,
    boxrule=0mm,
    breakable,
    #1
}

\newtcolorbox{PromptBox}[1][]{
    colback=lime!5!white,
    colframe=lime!75!black,
    title=\textbf{Model},
    top=2mm,
    bottom=2mm,
    left=2mm,
    right=2mm,
    arc=2mm,
    boxsep=1mm,
    boxrule=0mm,
    breakable,
    #1
}

\usepackage{newfloat}
\usepackage{listings}
\DeclareCaptionStyle{ruled}{labelfont=normalfont,labelsep=colon,strut=off} 
\floatstyle{ruled}
\newfloat{listing}{tb}{lst}{}
\floatname{listing}{Listing}
\title{DSCodeBench: A Realistic Benchmark for Data Science Code Generation}
\author {
    Shuyin Ouyang\textsuperscript{\rm 1}, 
    Dong Huang\textsuperscript{\rm 2},
    Jingwen Guo\textsuperscript{\rm 1},
    Zeyu Sun\textsuperscript{\rm 3},
    Qihao Zhu\textsuperscript{\rm 4},
    Jie M. Zhang\textsuperscript{\rm 1}
}
\affiliations {
    \textsuperscript{\rm 1}King's College London\\
    \textsuperscript{\rm 2}Institute of Data Science, National University of Singapore\\
    \textsuperscript{\rm 3}Institute of Software Chinese Academy of Sciences\\
    \textsuperscript{\rm 4}Peking University
}

\begin{document}

\maketitle

\begin{abstract}
We introduce DSCodeBench, a new benchmark designed to evaluate large language models (LLMs) on complicated and realistic data science code generation tasks.
DSCodeBench consists of 1,000 carefully constructed problems sourced from realistic problems from GitHub across ten widely used Python data science libraries.
DSCodeBench offers a more challenging and representative testbed, more complex code solutions, more comprehensive data science libraries, clearer and better structured problem descriptions, and stronger test suites.
To construct the DSCodeBench, we develop a robust pipeline that combines task scope selection, code construction, test case generation, and problem description synthesis.
The process is paired with rigorous manual editing to ensure alignment and enhance the reliability of the evaluation.
Experimental result shows that DSCodeBench exhibits robust scaling behavior, where larger models systematically outperform smaller ones, validating its ability to distinguish model capabilities.
The best LLM we test, GPT-4o, has a pass@1 of 0.392, indicating that LLMs still have a large room to improve for realistic data science code generation tasks. 
We believe DSCodeBench will serve as a rigorous and trustworthy foundation for advancing LLM-based data science programming.
\end{abstract}


\section{Introduction}
\label{section: Introduction}

Recent advances in large language models (LLMs) have significantly accelerated research in automated code generation, particularly for data science tasks that involve complex workflows and domain-specific libraries~\cite{bolyen2019reproducible, hassani2023role}.
As a result, a growing number of benchmarks have been proposed to evaluate LLMs in this setting, including DS-1000~\cite{lai2023ds}, DA-Code~\cite{huang2024code}, DataSciBench~\cite{zhang2025datascibench}, and others.
Among them, DS-1000 has emerged as a standard for evaluating LLMs on data science code generation.
However, it still falls short in capturing the full complexity and realism of real-world data science coding scenarios.
We identify three key limitations of DS-1000 that motivate the need for a more robust and realistic benchmark:
(1) Lack of realistic reference code.
Most code problems are from Stack Overflow.
Their solutions are usually easy, one-off code snippets, which cannot represent the complex scenarios encountered in real-world programming.
For example, the average length of reference code in DS-1000 is only 3.6 lines, which limits its ability to reflect realistic programming tasks.
(2) Insufficient testing.
DS-1000 includes, on average, fewer than 2.1 tests for each coding problem~\cite{lai2023ds}, which leads to insufficient coverage of program behavior.
As a result, many test cases are relatively simple and fail to fully explore the code's functionality or corner cases.
(3) Poorly structured problem description.
Each code problem description in DS-1000 includes a natural language description and code context, but these elements are often inconsistently formatted, which can limit clarity regarding the expected program behaviors.
These limitations highlight the need for a more comprehensive and realistic benchmark for data science code generation.

To fill this gap, we develop a pipeline to systematically construct realistic data science code generation tasks from GitHub.
Using this pipeline, we build DSCodeBench, a benchmark with 1,000 problems covering ten widely-used Python data science libraries, namely, NumPy, Pandas, SciPy, Scikit-learn, TensorFlow, PyTorch, Matplotlib, Seaborn, Keras, and LightGBM.
DSCodeBench offers a more challenging and representative testbed, featuring longer solution code (averaging 22.5 vs. 3.6 lines in DS-1000) and richer, better structured problem descriptions (averaging 474 vs. 140 words in DS-1000). 
It also provides much stronger tests (averaging 200 vs. 2.1 tests in DS-1000) with an evaluation framework supporting customizable test case configurations.

We evaluate ten state-of-the-art LLMs on DSCodeBench, and the results highlight DSCodeBench's challenging nature and evaluation reliability.
In particular, the best performing LLM, GPT-4o, has pass@1 score of 0.392.
Open-source coding-specific models, including DeepSeek-Coder and Qwen2.5-Coder variants, have a pass@1 score of 0.222 and 0.229, respectively.
We also observe a clear trend of scaling behavior, where larger models yield higher performance,  a trend not always evident on DS-1000. 
Together, these findings demonstrate that DSCodeBench serves as a more challenging, rigorous, realistic, and trustworthy benchmark for evaluating and advancing LLM-based data science code generation.

To conclude, this paper makes the following contributions.

\begin{itemize}[leftmargin=1cm] 
    \item We introduce DSCodeBench, a realistic benchmark designed to evaluate LLM performance on complicated data science code generation tasks.
    \item We develop a pipeline that constructs the benchmark from GitHub repositories, including task scope determination, code construction, test case generation, problem description synthesis, and manual editing.
    \item We perform a comprehensive empirical evaluation of 10 state-of-the-art LLMs on DSCodeBench. The benchmark, code, and experiment results are available at https://github.com/ShuyinOuyang/DSCodeBench.
\end{itemize}

\section{DSCodeBench Construction}
\label{section: DSCodeBench Construction}

\begin{figure*}
\centerline{\includegraphics[width=0.9\linewidth]{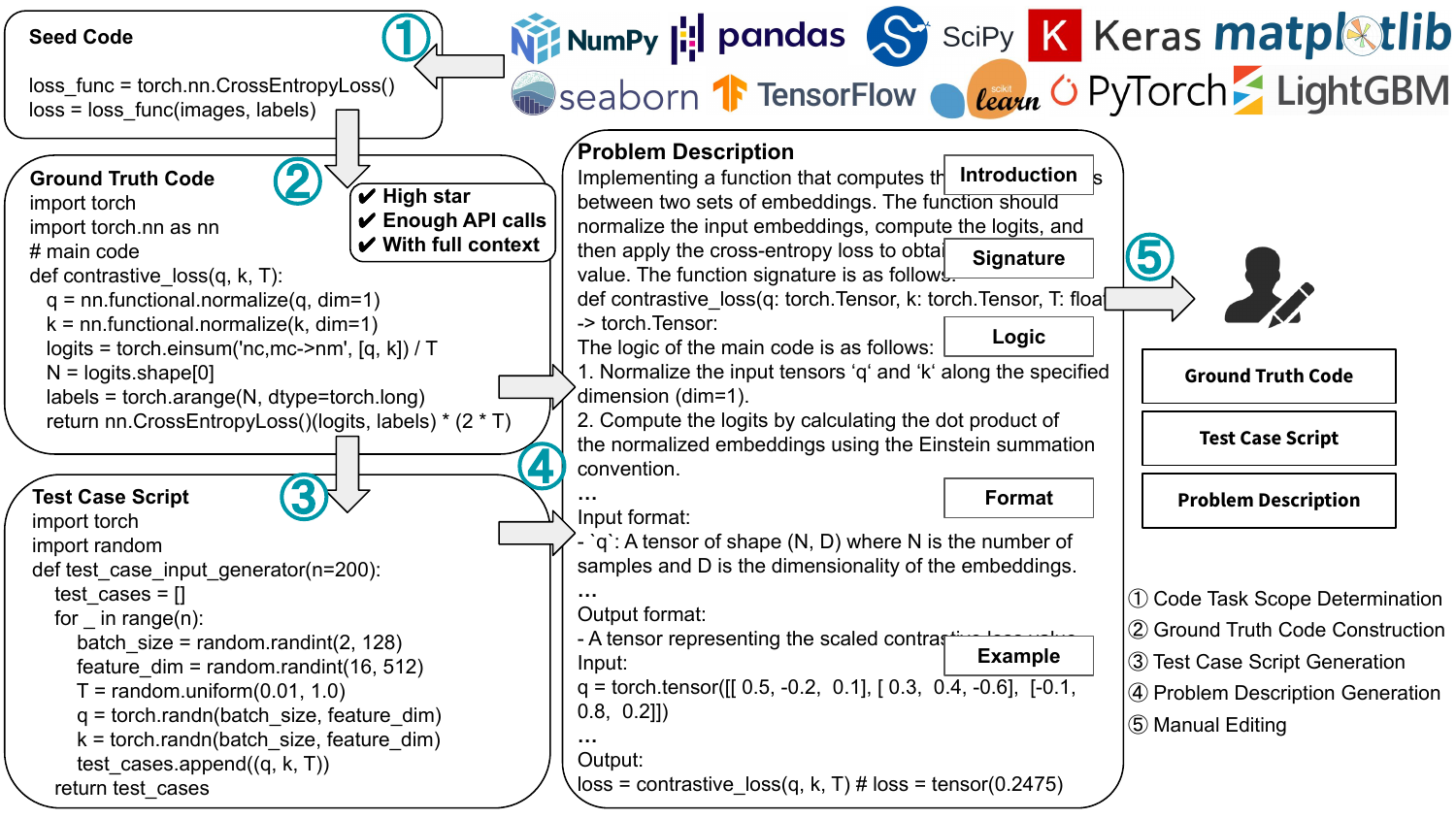}}
\vspace{0mm}
\caption{An example of DSCodeBench construction.
The pipeline begins by determining the task scope, followed by collecting seed code and constructing ground truth code.
This ground truth code is then used to generate a corresponding test case script that produces tailored input–output examples.
Using both the ground truth and the generated test cases, the problem description is generated.
Finally, all components are manually reviewed and aligned.}
\label{fig: benchmark construction example}
\end{figure*}

To ensure both the scale and quality of DSCodeBench, we adopt a structured pipeline consisting of task scope determination, ground truth code selection, test case generation, problem description generation, and manual editing.
We first determine the scope to specify the categories of programming tasks.
Next, we construct ground truth code from GitHub, which serves as the basis for subsequent steps.
We then generate a test case script from the ground truth code, which is used to automatically produce test cases for evaluation.
Problem descriptions are subsequently created to describe each task.
Finally, systematic manual editing is performed to ensure consistency and correctness across all components.
Figure~\ref{fig: benchmark construction example} provides an illustrative overview of our benchmark construction process.
The following sections introduce the details of each step in this process.

\subsection{Code Task Scope Determination}
\label{subsection: Code Task Scope Determination}

The construction of the DSCodeBench begins with the careful choosing of the code task scope.
Our goal is to ensure comprehensive coverage of widely used data science libraries, thereby capturing the practical challenges commonly encountered by developers.
We extend the foundation laid by the existing DS-1000 benchmark, which covers seven libraries, by expanding the set to ten. 
In particular, we add Seaborn, Keras, and LightGBM to better capture the evolving landscape of data science.
These libraries have seen significant adoption in recent years: Seaborn for its advanced statistical visualization capabilities, Keras as a widely used high-level interface for deep learning, and LightGBM for its scalable implementation of gradient boosting algorithms.
Their inclusion addresses the increased demand for code generation benchmarks involving complex APIs and diverse usage patterns that pose additional challenges for LLMs.
This broader coverage increases the diversity and practical relevance of DSCodeBench, establishing it as a more representative and robust benchmark for assessing code generation models in real-world data science workflows.

\subsection{Ground Truth Code Construction}
\label{subsection: Ground Truth Code Construction}

The first stage focuses on constructing high-quality ground truth code to create more realistic code generation tasks, consisting of four sub-steps: collecting seed code, sourcing code from GitHub, reconstructing missing context, and filtering the code candidates.

\paragraph{Collecting seed code.}
We begin by collecting seed code that serves as the query for retrieving realistic data science code from GitHub. 
The code is obtained from two sources: the reference code in the DS-1000 benchmark, and the answer code extracted from Stack Overflow coding questions.
Inspired by DS-1000, we use library names such as “numpy” and “pandas” as keywords to search for highly voted Stack Overflow questions, collecting up to 500 top-ranked questions per library.
For each selected question, we extract the highest-voted answer's code as seed code.

\paragraph{Sourcing code from GitHub.} 
Each snippet is then split at the line level, and each line is used as a query to retrieve code from GitHub via the GitHub REST API.
We filter these results to retain only Python files and apply deduplication to remove redundant candidates, resulting in a pool of 807,198 code candidates, each stored with its associated repository metadata for subsequent processing.

\paragraph{Context reconstruction.}
We perform context reconstruction for two main reasons. 
First, the retrieved code candidates are often snippets that lack the surrounding context necessary for standalone execution. 
Second, using raw GitHub code directly poses a risk of data leakage, as large language models may have been pre-trained on the same public code, potentially leading to unfair evaluation.
To address both challenges, we design an automated context reconstruction pipeline that transforms code candidates into standardized, self-contained units. 
For each candidate snippet, we locate its original source file within the corresponding repository and parse it into an Abstract Syntax Tree (AST) for structural analysis. 
We then perform dependency resolution, identifying and extracting all required contextual elements, such as import statements, variable initializations, and auxiliary function definitions.
In cases where the candidate represents a method encapsulated within a class, we refactor it into a top-level function, transforming its object-oriented dependencies into explicit parameters as needed.
To ensure functional completeness, if any referenced variables or functions remain unresolved after context reconstruction, we promote them to explicit function parameters, enabling the candidate to be executed in isolation as a self-contained unit.
In addition, we apply code reformatting to further standardize the reconstructed candidates.
For instance, if the main code lacks a return statement, we append a return for the last variable (excluding those defined as function parameters). 
We systematically remove file-related operations to avoid external dependencies, which can mitigate the risk of data leakage.

\paragraph{Code filtering.}
We apply a two-stage filtering process, property and functional filtering, to ensure that only high-quality candidates are retained.
At the property level, we implement: 
(1) compilation filtering to eliminate candidates that fail to execute at the file-level independently; 
(2) star filtering to retain only candidates from repositories with at least 10 GitHub stars, thereby promoting code quality; and 
(3) API call filtering to ensure that each candidate contains at least three API calls, ensuring adequate task complexity.
At the functional level, we evaluate candidates based on their ability to successfully execute generated test cases.
For each candidate, we prompt the LLM to generate 10 diverse input test cases based on the code. 
A candidate is accepted if it executes successfully on at least one test case.
If all initial test cases fail, additional sets are iteratively generated until the candidate passes at least one or a predefined maximum of five attempts are reached.
Candidates that fail to pass any test case within this limit are discarded.
This process serves as a coarse-grained filter, designed to eliminate candidates with fundamental errors in the code logic.
In the subsequent stage, we employ a more stringent procedure to further verify the quality of the ground truth code.
Ultimately, this filtering pipeline yields 7,623 validated candidates, each paired with corresponding sample inputs and outputs, collectively constituting the ground truth code set for DSCodeBench.

\subsection{Test Case Generation}
\label{subsection: Test Case Generation}

Generating a large number of high-quality test cases for code problems is typically labor-intensive and costly.
To address this challenge, we design an automatic approach that generates test-case input generation scripts rather than manually crafting individual test cases.
We leverage LLMs to produce scripts that, when executed, are capable of generating diverse and valid test inputs for the corresponding ground truth code.
However, not every generated script produces fully correct test cases.
In particular, some generated test inputs may fail when passed to the ground truth code, raising runtime errors or exceptions.
To mitigate this issue and ensure the reliability of our test case generation process, we incorporate a script repair mechanism based on self-repair~\cite{xia2023keep}.

\paragraph{Test Case Script Improvement.}
We employ a self-repair loop to iteratively enhance the quality of the generated test case scripts.
Assuming the ground truth code obtained from the previous stage is correct, we enforce a strict criterion: all generated test case inputs must execute successfully without any errors, whether arising from the input data or the code logic.
If a generated input triggers an error, we record both the input and the corresponding error message.
This information is then used to guide the LLM in revising the test case script to eliminate the identified issues.
The repair process is repeated for up to five iterations or until the script reliably generates valid inputs that pass all tests.
This iterative refinement improves the robustness and reliability of the test case scripts, ensuring they satisfy the stringent standards of DSCodeBench.
To further assess the effectiveness of the generated test cases, we perform a test coverage analysis by applying them to the ground truth code. 
Following this step, 2,407 code candidates are retained.

\subsection{Problem Description Generation}
\label{subsection: Problem Description Generation}

Given the ground truth code, we leverage LLM to generate corresponding code problem descriptions, aiming to produce better-structured and well-formatted task descriptions.
Each description is required to include at least five components: a general introduction, a function signature, logic, input and output formats, and illustrative input-output examples.
However, we observe that LLMs often struggle to generate accurate and consistent input-output pairs.
Although examples may appear plausible, they frequently fail to match the ground truth code's behavior, thereby compromising benchmark quality.
To address this issue, we divide the description generation process into two separate stages.

First, we prompt the LLM to generate a problem description covering only the general introduction, the function signature, logic, and the input-output format, explicitly excluding examples.
Second, we generate an input-output example independently by executing the test case generation scripts produced in the previous step.
Finally, we concatenate the LLM-generated description and the independently generated example to form the final code problem description.
This separation ensures that examples remain consistent with the underlying code logic.
However, misalignments between the generated description and the ground truth code may still occur.
To further ensure alignment quality, we perform manual editing in the next step.

\subsection{Manual Editing}
\label{subsection: Manual Editing}

After completing the automated stages of construction, we obtain a preliminary pool of 1,000 code problem sets, each containing a ground truth code snippet, its corresponding problem description, and a test case generation script.
To ensure the benchmark quality, we conduct a rigorous manual review and refinement process.
Manual editing is collaboratively performed by four authors with expertise in programming and data science.

First, we double-check the test case generation scripts to ensure their robustness and variability.
Specifically, we modify the random seeds used in the scripts and verify that they consistently generate valid and diverse test cases across different runs, reinforcing the reliability of the benchmark.
Second, we validate the alignment between the ground truth code, the test case script, and the problem description.
Specifically, we ensure that the ground-truth code aligns with the input-output examples in the problem description through execution-based validation. 
We also verify that the core logic outlined in the problem description accurately reflects the ground-truth implementation and provides sufficient information for a human to solve the data science task, using both human judgment and LLM-as-a-judge~\cite{gu2024survey} for evaluation. 

To mitigate the risk of data leakage, we do not directly reuse code extracted from GitHub repositories. Instead, we apply systematic perturbations to the extracted code to produce semantically similar solutions.
These transformations include modifying function signatures (e.g., changing parameter names and return types), adding or removing lines of code while preserving the overall functionality, and restructuring control flows based on the code’s context.
While it is possible that similar code may have been seen during pretraining, our benchmark provides human-written problem descriptions that differ significantly from any original code comments or documentation, reducing the likelihood of overlap in both code and natural language.
We conduct a similarity analysis between LLM-generated code and ground truth code. The text similarity (less than 0.4) and AST similarity (less than 0.5) confirm the effectiveness.

Furthermore, DSCodeBench is designed to support flexible evaluation settings.
The benchmark provides users with the ability to customize the test cases by adjusting the random seed and specifying the number of test cases required, enabling more adaptable and robust evaluation protocols.

\section{Benchmark Statistics}
\label{section: Benchmark Statistics}

\begin{figure}
\centerline{\includegraphics[width=\linewidth]{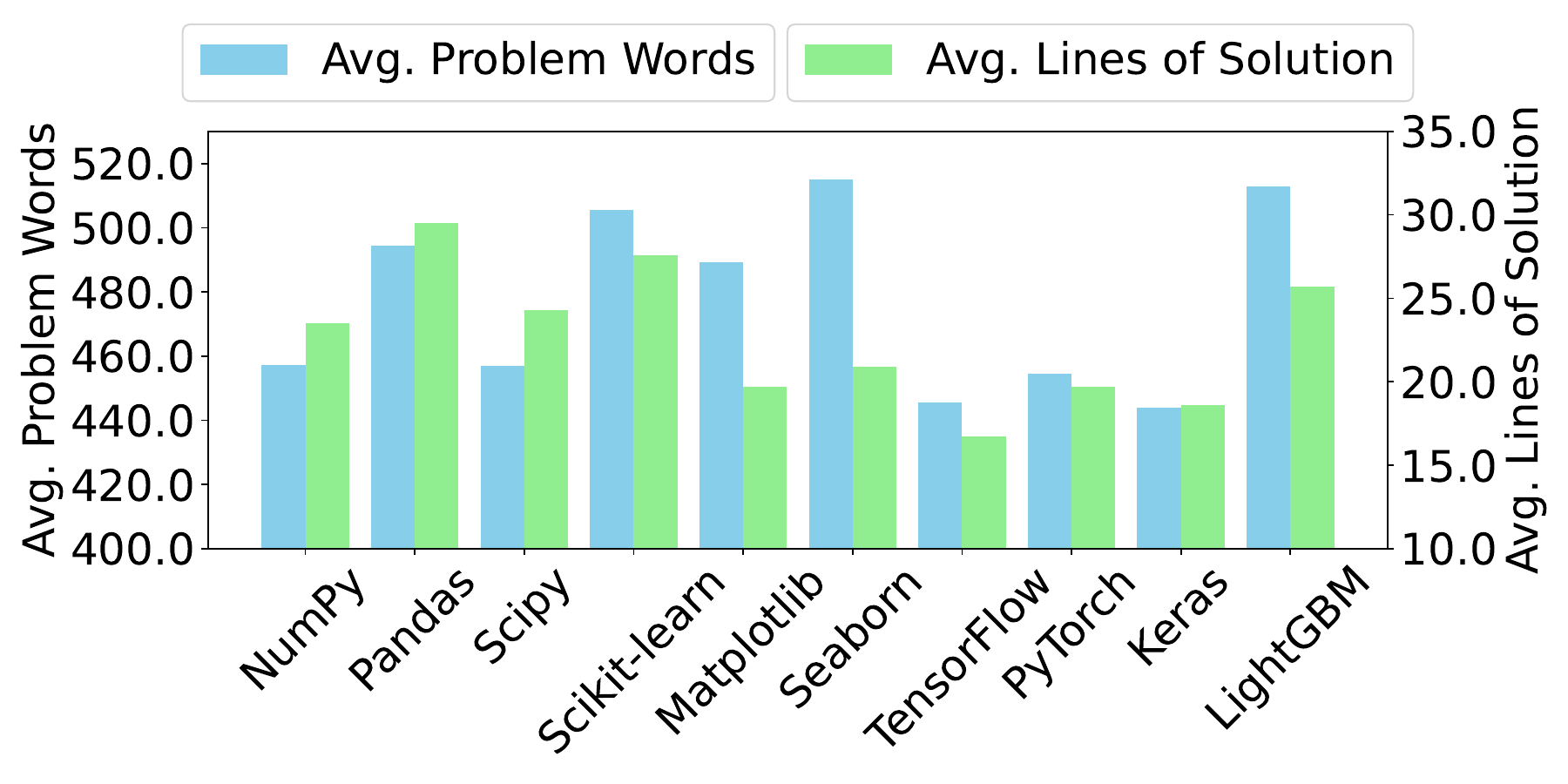}}
\vspace{0mm}
\caption{Distribution of tasks in DSCodeBench by library.}
\label{fig: DSCodeBench distribution}
\end{figure}

DSCodeBench focuses on ten data science libraries, including 131 problems for NumPy, 92 for Pandas, 112 for SciPy, 108 for Scikit-learn, 105 for Matplotlib, 83 for Seaborn, 110 for TensorFlow, 101 for PyTorch, 104 for Keras, and 54 for LightGBM.
As shown in Figure~\ref{fig: DSCodeBench distribution}, the average number of problem words for different libraries stays between 443.8 and 514.9, and the average line of solutions stays between 16.7 and 29.5,
which demonstrate that tasks are not limited to short or simplified descriptions but instead reflect rich problem statements that mirror real-world project specifications.

We compared DSCodeBench with other code generation benchmarks. 
Table~\ref{table: DSCodeBench comparison} offers a comprehensive comparison between DSCodeBench and several widely used benchmarks for code generation, including HumanEval~\cite{chen2021evaluating}, MBPP~\cite{austin2021program}, APPS~\cite{hendrycks2021measuring}, BigCodeBench~\cite{zhuo2024bigcodebench}, LiveCodeBench~\cite{jain2024livecodebench}, DSP~\cite{chandel2022training}, DA-Code~\cite{huang2024code}, DataSciBench~\cite{zhang2025datascibench}, and DS-1000~\cite{lai2023ds}. 
When compared to general code generation benchmarks such as HumanEval, MBPP, APPS, BigCodeBench, and LiveCodeBench, DSCodeBench demonstrates significantly greater complexity and evaluation rigor, featuring more test cases per task and more intricate problem descriptions.
HumanEval and MBPP, while foundational, are limited by their concise problem descriptions (averaging 23.0 and 15.7 words, respectively) and short canonical solutions (6.3 and 6.7 lines on average), reflecting small, well-scoped programming tasks.
Even BigCodeBench and LiveCodeBench, which feature more challenging problems, offer a considerably lower evaluation depth with only 5.6 and 17.0 test cases per problem.
In contrast, DSCodeBench is detailed in problem description, averaging 474.0 words, and requires more extensive solutions (22.5 lines on average), while each task is evaluated with 200 test cases by default.

\begin{table*}[ht]
\caption{Comparison of DSCodeBench to other benchmarks. * refers to multiple solutions needed in one task. The top five are general code generation benchmarks, while the bottom four are data science–specific code generation benchmarks.}

\vspace{0mm}

\centering
\resizebox{\linewidth}{!}{
\begin{tabular}{l r r r r r r}
\toprule
Benchmark & No. of Problems & No. of Libraries & Avg. Test Cases & Avg. Problem Words & Avg. Lines of Solution & Data Source \\
\midrule
HumanEval & 164 & - & 7.1 & 23.0 & 6.3 & Hand-Written \\
MBPP & 974 & - & 3.0 & 15.7 & 6.7 & Hand-Written \\
APPS & 10,000 & - & 13.2 & 293.2 & 18.0 & Hand-Written \\
BigCodeBench & 1,140 & 62 & 5.6 & 147.8 & 10.0 & StackOverflow\\
LiveCodeBench & 511 & - & 17.0 & 278.0 & 13.6 & Online Coding Platforms \\
\midrule
DSP & 1119 & 8 & 2.1 & 71.9 & 4.5 & Notebooks \\ 
DA-Code & 500 & - & - & 40.2 & 85.0* & Annotation\\
DataSciBench & 222 & 6 & 2.3 & 130.5 & 24.7 & Multi-Sourced \\
DS-1000 & 1000 & 7 & 1.6 & 140.0 & 3.6 & StackOverflow \\
\midrule
\textbf{DSCodeBench} & 1000 & 10 & 200.0 (default) & 474.0 & 22.5 & GitHub \\
\bottomrule
\end{tabular}
}
\label{table: DSCodeBench comparison}
\end{table*}




Compared to existing data science code generation benchmarks, DSCodeBench further raises the standard across many dimensions.
Although DSP and DS-1000 are early efforts to adapt benchmarks toward the data science domain, they still maintain short problem descriptions (71.9 and 140.0 words on average) and short solutions (4.5 and 3.6 lines on average), and often focus on isolated code snippets extracted from notebooks or Q\&A forums.
Moreover, evaluation in these datasets remains insufficient, with only 2.1 and 1.6 test cases per problem on average. 
DA-Code and DataSciBench primarily focus on the diversity of data science tasks (e.g., data processing, visualization, and etc.), with tasks curated from real-world datasets. 
However, they provide insufficient test cases, particularly lacking comprehensive testing for intermediate generated code.
In contrast, DSCodeBench is built from GitHub repositories and emphasizes robust evaluation, ensuring that each problem represents a substantial, end-to-end data science task with robust evaluation test suites rather than an isolated code snippet.
It places particular focus on assessing whether the complex logic of the code is correctly implemented within the context of real-world coding tasks.

\section{Benchmark State-of-the-art LLMs}
\label{section: Benchmark State-of-the-art LLMs}

\subsection{Experiment Setup}
\label{subsection: Experiment Setup}

\paragraph{Models.} 
We evaluate both open-source and closed-source LLMs in code generation. 
For the closed-source model, we evaluate DSCodeBench with OpenAI's GPT models.
Specifically, we experiment with GPT-3.5-turbo, GPT-4o-mini, and GPT-4o. 
For the open-source model, the LLMs we used include DeepSeek-Coder-Instruct (1.3B, 6.7B, and 33B), DeepSeek-Coder-V2-Lite-Instruct (15.7B), and Qwen2.5-Coder-Instruct (7B, 14B, and 32B).

\paragraph{Randomness control.} LLMs are non-deterministic in nature~\cite{ouyang2025empirical}. 
We set the temperature of all models to 0.2, while keeping all other parameters at their default values.
We also run all the models 3 times identically to mitigate the randomness that affects our experiment results. 
For each run, we collect the code snippets from the LLM's response, evaluate them, and record their test case passing status, whether they fully pass the test cases, partially pass the test cases, or fail all the test cases.

\paragraph{Compute resource.} The experiments are conducted in an edge server with an Intel Xeon Platinum 8336C CPU
with 128 cores, 8 * NVIDIA A100-SXM GPUs, and a total memory capacity of 2.0TiB.

\subsection{Metrics} 
\label{subsection: Metrics}


\paragraph{Pass@k} We measure the performance with pass@k~\cite{chen2021evaluating}, where k code samples are generated per problem. 
A problem is considered solved if any sample passes the unit tests. 
The fraction of problems solved is reported.
We pick k as 1 and 3 in our experiments.
We run the experiments 3 times, and report the mean pass@1 and pass@3 in our results.

\paragraph{Test case passing status} In this paper, we classify the passing status into three categories, namely correct, partially correct, and wrong.
Correct means that the generated code can pass all the test cases.
Partially correct means that the code can only pass part of all the test cases.
Wrong means that the code fails to pass any test cases.
We record the average numbers for each category and use them to reflect the code generation performance.

\begin{table*}[ht]\small
\caption{Experiment results on DSCodeBench and DS-1000.}
\vspace{0mm}
\centering
\resizebox{\linewidth}{!}{
\begin{tabular}{l l r r r r r}
\toprule
Benchmark & Model & Pass@1 & Pass@3 & Avg. Correct & Avg. Partially Correct & Avg. Wrong \\
\midrule
\multirow{10}{*}{DS-1000} & GPT-3.5-turbo & 0.374 & 0.442 & 373.7$\pm$6.2 & 0.0$\pm$0.0 & 626.3$\pm$6.2 \\
& GPT-4o-mini & 0.422 & 0.485 & 422.0$\pm$8.6 & 0.0$\pm$0.0 & 578.0$\pm$8.6 \\
& GPT-4o & 0.451 & 0.545 & 450.7$\pm$2.9 & 0.0$\pm$0.0 & 549.3$\pm$2.9 \\
\cmidrule{2-7}
& DeepSeek-Coder-1.3B-Instruct & 0.122 & 0.230 & 121.7$\pm$4.1 & 0.0$\pm$0.0 & 878.3$\pm$4.1 \\
& DeepSeek-Coder-6.7B-Instruct & 0.172 & 0.301 & 172.0$\pm$4.2 & 0.0$\pm$0.0 & 828.0$\pm$4.2 \\
& DeepSeek-Coder-V2-Lite-Instruct(15.7B) & 0.228 & 0.337 & 228.0$\pm$8.5 & 0.0$\pm$0.0 & 772.0$\pm$8.5 \\
& DeepSeek-Coder-33B-Instruct & 0.136 & 0.241 & 136.0$\pm$3.7 & 0.0$\pm$0.0 & 864.0$\pm$3.7 \\
\cmidrule{2-7}
& Qwen2.5-Coder-7B-Instruct & 0.005 & 0.012 & 4.7$\pm$2.5 & 0.0$\pm$0.0 & 995.3$\pm$2.5 \\
& Qwen2.5-Coder-14B-Instruct & 0.323 & 0.476 & 323.0$\pm$10.2 & 0.0$\pm$0.0 & 677.0$\pm$10.2 \\
& Qwen2.5-Coder-32B-Instruct & 0.419 & 0.508 & 419.0$\pm$3.7 & 0.0$\pm$0.0 & 581.0$\pm$3.7 \\

\midrule
\midrule
\multirow{10}{*}{DSCodeBench} & GPT-3.5-turbo & 0.307 & 0.346 & 307.3$\pm$5.6 & 82.3$\pm$3.4 & 610.3$\pm$6.3 \\
& GPT-4o-mini & 0.342 & 0.374 & 342.3$\pm$1.2 & 77.3$\pm$3.3 & 580.3$\pm$4.5 \\
& GPT-4o & 0.392 & 0.438 & 391.7$\pm$4.6 & 105.0$\pm$4.3 & 503.3$\pm$0.5 \\
\cmidrule{2-7}
& DeepSeek-Coder-1.3B-Instruct & 0.076 & 0.103 & 76.3$\pm$3.3 & 27.0$\pm$2.2 & 896.7$\pm$5.4 \\
& DeepSeek-Coder-6.7B-Instruct & 0.163 & 0.195 & 162.7$\pm$2.1 & 47.0$\pm$1.6 & 790.3$\pm$2.1 \\
& DeepSeek-Coder-V2-Lite-Instruct(15.7B) & 0.205 & 0.234 & 205.0$\pm$1.4 & 48.0$\pm$0.8 & 747.0$\pm$0.8 \\
& DeepSeek-Coder-33B-Instruct & 0.222 & 0.258 & 222.3$\pm$2.6 & 51.0$\pm$5.4 & 726.7$\pm$4.0 \\
\cmidrule{2-7}
& Qwen2.5-Coder-7B-Instruct & 0.116 & 0.164 & 116.3$\pm$7.1 & 28.7$\pm$0.5 & 855.0$\pm$7.5 \\
& Qwen2.5-Coder-14B-Instruct & 0.213 & 0.251 & 212.7$\pm$5.6 & 49.0$\pm$1.4 & 738.3$\pm$5.3 \\
& Qwen2.5-Coder-32B-Instruct & 0.229 & 0.260 & 228.7$\pm$2.6 & 45.7$\pm$4.2 & 725.7$\pm$2.5 \\

\bottomrule
\end{tabular}
}
\label{table: DSCodeBench experiment mean+std}
\end{table*}

\subsection{Experiment Results}
\label{subsection: Experiment Results}

\paragraph{Closed-source models}
As shown in Table~\ref{table: DSCodeBench experiment mean+std}, closed-source models such as GPT-3.5-turbo, GPT-4o-mini, and GPT-4o achieve the highest overall performance on DSCodeBench.
Among them, GPT-4o achieves the highest pass@1 score of 0.392, the highest pass@3 score of 0.438, and the largest average number of correct outputs (391.7) across all evaluated models.
Notably, the smaller GPT-4o-mini also surpasses GPT-3.5-turbo in all key metrics, indicating that architectural and training optimizations play a critical role beyond model scaling alone.
Despite these advances, \textbf{the relatively low pass@1 scores of even the best models reflect the challenge posed by DSCodeBench in realistic and diverse data science code generation tasks}.

\paragraph{Open-source models}
In contrast, open-source code generation models, including DeepSeek and Qwen variants, lag significantly behind closed-source models on DSCodeBench.
The strongest open-source model, DeepSeek-Coder-33B-Instruct, reaches pass@1 of only 0.222, which remains substantially below GPT-4o’s 0.392.
Similarly, the average numbers of fully correct and partially correct outputs are consistently lower, while the number of wrong outputs remains higher.
However, the clear and consistent performance gains observed with increasing model sizes and improved versions indicate that \textbf{our chosen models on DSCodeBench follow the scaling law}.

\paragraph{Comparison with DS-1000} 
Compared to DS-1000, models consistently achieve lower pass@1 and pass@3 scores on DSCodeBench.
For example, GPT-4o obtains pass@1 and pass@3 of 0.451 and 0.545 on DS-1000 but only 0.392 and 0.438 on DSCodeBench.
This decline highlights that (1) DSCodeBench presents a more challenging task for LLMs.
In addition, the variance of the evaluation metrics, such as the average number of correct outputs, is relatively smaller in DSCodeBench than in DS-1000. 
This lower variance reflects that (2) DSCodeBench provides a more consistent and reliable test suite, where evaluation outcomes are less affected by randomness and better reflect the capabilities of models. 
Furthermore, (3) the results on DSCodeBench exhibit clear adherence to the scaling law, with larger models systematically outperforming smaller ones.
Within both DeepSeek and Qwen model families, increasing the size of parameters consistently leads to higher pass@1 and pass@3 scores and better correctness.
For example, in the Qwen2.5-Coder series, the pass@1 steadily increases from 0.116 (7B) to 0.213 (14B) and 0.229 (32B), reflecting the critical role of scaling even within the same architectural family.
In contrast, DS-1000 shows irregular scaling behavior. 
\textbf{The results show that DSCodeBench not only increases task difficulty but also offers a more rigorous, stable, and trustworthy benchmark to assess real-world data science code generation.}

Looking deeper into the benchmark, we observe that DS-1000 exhibits several issues that undermine its reliability as a benchmark, particularly in cases where it fails to reflect expected scaling behavior.
First, the code context in DS-1000’s problem descriptions makes it difficult for models to generate valid solutions that conform to its expected format, which is designed for code completion and insertion.
As a result, models often produce either empty outputs or solutions that fail to meet basic structural requirements.
Second, the reference solutions in DS-1000 are typically very short, and models tend to mimic this brevity, often resulting in repeated or incomplete outputs that encounter the response length constraint.
Third, the evaluation is often based on a small set of test cases, which contributes to higher variance and unstable model performance.
In contrast, DSCodeBench addresses these challenges in three aspects.
First, its problem descriptions are designed to be aligned with ground truth code and standardized, reducing ambiguity and helping models generate structurally valid solutions. 
Second, the solutions are of realistic length, avoiding unintended biases toward brevity and minimizing the risk of incomplete or constrained outputs.
Third, DSCodeBench employs a larger set of test cases, which leads to more reliable model performance.
Together, these improvements make DSCodeBench a more robust and reliable benchmark than DS-1000 for evaluating LLMs on data science code generation tasks.

\begin{table*}[ht]\tiny
\caption{Experiment result on DSCodeBench for each library (pass@1).}
\vspace{0mm}
\centering
\resizebox{\linewidth}{!}{
\begin{tabular}{l r r r r r r r r r r}
\toprule
Model & NumPy & Pandas & Scipy & Scikit-learn & Matplotlib & Seaborn & TensorFlow & PyTorch & Keras & LightGBM \\
\midrule
GPT-3.5-turbo & 0.226 & 0.239 & 0.315 & 0.330 & 0.311 & 0.149 & 0.400 & 0.498 & 0.324 & 0.216\\
GPT-4o-mini & 0.280 & 0.348 & 0.432 & 0.435 & 0.238 & 0.124 & 0.373 & 0.488 & 0.356 & 0.290\\
GPT-4o & 0.405 & 0.366 & 0.482 & 0.485 & 0.210 & 0.141 & 0.452 & 0.591 & 0.385 & 0.290\\
\midrule
DeepSeek-Coder-1.3B-Instruct & 0.097 & 0.036 & 0.000 & 0.176 & 0.022 & 0.060 & 0.115 & 0.142 & 0.042 & 0.049\\
DeepSeek-Coder-6.7B-Instruct & 0.193 & 0.083 & 0.027 & 0.340 & 0.029 & 0.076 & 0.300 & 0.330 & 0.096 & 0.080\\
DeepSeek-Coder-V2-Lite-Instruct(15.7B) & 0.193 & 0.149 & 0.027 & 0.401 & 0.013 & 0.129 & 0.333 & 0.465 & 0.179 & 0.099\\
DeepSeek-Coder-33B-Instruct & 0.282 & 0.149 & 0.027 & 0.432 & 0.029 & 0.096 & 0.373 & 0.479 & 0.141 & 0.130\\
\midrule
Qwen2.5-Coder-7B-Instruct & 0.158 & 0.029 & 0.018 & 0.201 & 0.010 & 0.052 & 0.239 & 0.244 & 0.106 & 0.037\\
Qwen2.5-Coder-14B-Instruct & 0.275 & 0.156 & 0.036 & 0.327 & 0.029 & 0.177 & 0.367 & 0.396 & 0.196 & 0.086\\
Qwen2.5-Coder-32B-Instruct & 0.336 & 0.174 & 0.036 & 0.299 & 0.022 & 0.149 & 0.430 & 0.449 & 0.173 & 0.130\\
\bottomrule
\end{tabular}
}
\label{table: DSCodeBench experiment each library(pass@k)}
\end{table*}

\paragraph{Library-level analysis} Table~\ref{table: DSCodeBench experiment each library(pass@k)} presents a detailed breakdown of the mean pass@1 scores on DSCodeBench across ten libraries.
Closed-source models, particularly GPT-4o, dominate performance in nearly all libraries.
GPT-4o consistently achieves the highest pass@1 scores (e.g., 0.405 for NumPy, 0.482 for SciPy, and 0.591 for PyTorch).
In contrast, open-source models exhibit lower overall performance but still show reasonable scaling trends.
Within both the DeepSeek and Qwen families, larger models outperform smaller ones across most libraries. 
For instance, DeepSeek-Coder-33B-Instruct shows clear improvements over DeepSeek-Coder-1.3B-Instruct, especially in libraries like scikit-learn (0.432) and PyTorch (0.479).
Similarly, Qwen2.5-Coder-32B-Instruct shows notable gains compared to its smaller variants.
Beyond cases where the generated logic fails to satisfy the task requirements, common errors include data structure mismatches, such as incompatible shapes or types during transformation steps, particularly in NumPy, TensorFlow, and PyTorch tasks.
Visualization libraries also exhibit frequent issues, with models often mis-specifying figures or plotting arguments.

However, certain libraries, particularly Matplotlib and Seaborn, remain challenging across all models.
The pass@1 scores for these libraries are consistently lower than for libraries like Scikit-learn, Keras, or PyTorch.
This suggests that generating data visualization remains difficult for current LLMs, particularly under our evaluation criterion requiring a similarity score exceeding 50\%.
Even DeepSeek-Coder-33B-Instruct only achieves 0.029 and 0.096 pass@1 on Matplotlib and Seaborn tasks, respectively.
The results reveal that while current models are increasingly capable of handling core data science libraries like NumPy and PyTorch, significant challenges remain for specialized domains such as visualization.
\textbf{The detailed breakdown shows that DSCodeBench successfully covers a broad and challenging spectrum of real-world data science coding tasks.}

\section{Related Work}
\label{section: Related Work}

\subsection{LLM4Code}
\label{subsection: LLM4Code}

LLMs and their agent frameworks have advanced code intelligence demonstrating strong capabilities in code generation, completion, translation, repair, and summarization~\cite{chen2021evaluating, macedo2024intertrans, yang2024revisiting, ahmed2024automatic, huang2023agentcoder, huang2023codecot}. 
Recent studies explore scaling laws and pretraining objectives for code-specific LLMs~\cite{hui2024qwen2, guo2024deepseek}, semantic integration for improved understanding~\cite{macedo2024intertrans}, and retrieval-augmented or prompt-based methods to tackle complex coding tasks~\cite{ouyang2025empirical, tan2024prompt, tao2024magis, huang2024effi, huang2024effilearner}.
Applications include test generation~\cite{ryan2024code,huang2024rethinking} and program repair~\cite{chen2024code}.
Despite this progress, challenges remain in addressing hallucination, and the need for domain-specific tuning~\cite{gu2025effectiveness, ouyang2025knowledge}, underscoring the importance of task-oriented benchmarks with robust evaluation.

\subsection{Code Generation Benchmark}
\label{subsection: Code Generation Benchmark}

With growing model capabilities, benchmarks for code generation \cite{chen2021evaluating, huang2024bias, huang2024effibench, qing2025effibenchx, hu2025dynacode, zhuo2024bigcodebench, jain2024livecodebench, huang2024code, zhang2025datascibench, chen2025dycodeeval, chen2024ppm, xiang2025scireplicate} have evolved in difficulty and scope.
HumanEval~\cite{chen2021evaluating} set a standard for evaluating functional correctness from natural language.
Subsequent datasets expanded this, including APPS~\cite{hendrycks2021measuring} for diverse coding tasks, MBPP~\cite{austin2021program} for beginner Python problems, and CodeContests~\cite{li2022competition} for competitive programming.
Domain-specific benchmarks emerged with DS-1000~\cite{lai2023ds}, focusing on real-world data science tasks, and SWE-bench~\cite{jimenez2023swe}, emphasizing realistic software engineering workflows.
However, data science code benchmarks remain limited: many rely on synthetic code, vague task descriptions, and insufficient test cases (e.g., fewer than 2.3 tests per task~\cite{lai2023ds, zhang2025datascibench}).
To address this, we introduce DSCodeBench, a benchmark grounded in realistic data workflows, with well-formatted descriptions and comprehensive test suites, offering a more rigorous evaluation framework for LLM-based data science code generation.

\section{Conclusion and Future Work}
\label{section: Conclusion and Future work}


In this work, we present DSCodeBench, a benchmark dataset designed to evaluate the performance of LLMs on realistic data science code generation tasks.
DSCodeBench consists of 1,000 problems constructed from real-world use cases across ten widely used Python data science libraries.
In contrast to prior benchmarks that often focus on simplified or artificial tasks with limited evaluation rigor, DSCodeBench provides more challenging, library-specific problems alongside a robust evaluation framework with comprehensive test cases.
By targeting library-aware code generation, DSCodeBench aims to bridge the gap between benchmark tasks and real-world programming scenarios.

Building on this foundation, several directions remain for future development. 
First, extending beyond Python to include languages like R would enable broader evaluation across diverse data science ecosystems.
Second, incorporating more complex code structures, such as error handling, multi-function logic, and project-level tasks, would better reflect real-world workflows.
Third, future versions could go beyond functional correctness by evaluating runtime performance, security, and adherence to coding best practices.
Lastly, integrating human or LLM-based judges to assess readability and style could complement automated testing for a more holistic evaluation.


\newpage
\section{Acknowledgement}
This work was supported by ITEA Genius
and ITEA GreenCode projects (funded by InnovateUK), the UKRI Centre for Doctoral Training in Safe and Trusted Artificial Intelligence (EP/S023356/1), and the National Natural Science Foundation of China (62402482).

\bibliography{reference}
\newpage
\definecolor{lightorange}{RGB}{253,188,180}
\definecolor{lightblue}{RGB}{180,208,253}
\definecolor{problemcolor}{RGB}{153,0,153}
\definecolor{modelcolor}{RGB}{0, 0, 255}

\definecolor{pythonblue}{RGB}{84, 184, 255}
\definecolor{soappink}{RGB}{255, 163, 203}




%
\lstset{%
	basicstyle={\footnotesize\ttfamily},
	numbers=left,numberstyle=\footnotesize,xleftmargin=2em,
	aboveskip=0pt,belowskip=0pt,%
	showstringspaces=false,tabsize=2,breaklines=true}
\floatstyle{ruled}
\newfloat{listing}{tb}{lst}{}
\floatname{listing}{Listing}
%
\pdfinfo{
/TemplateVersion (2026.1)
}

\setcounter{secnumdepth}{0} 

%




\section{Appendix}



\subsection{Limitation}
\label{appendix: Limitation}

While DSCodeBench significantly advances the evaluation of LLMs in data science code generation, there are still limitations.

First, DSCodeBench focuses exclusively on Python and ten popular data science libraries. 
Although these libraries cover a wide range of real-world data science workflows, DSCodeBench does not currently assess LLM capabilities across other programming languages. 
Extending coverage beyond Python remains an important direction for future work.

Second, although the code problems are drawn from real-world GitHub repositories, the extraction and curation process inherently involve some filtering and simplifications.
In particular, to improve the robustness of automatic test case generation, we simplify error-handling logic: error-raising code segments are either removed or replaced with default behaviors such as returning None.
While this choice enables more reliable automated evaluation, it slightly reduces the fidelity of error-related coding patterns in the dataset.
Furthermore, some highly complex or multi-file codebases are excluded, meaning that DSCodeBench primarily targets single-function or single-file tasks rather than full project-level development workflows.

Third, DSCodeBench evaluates functional correctness based on unit tests, without explicitly assessing other important dimensions of code quality, such as computational efficiency, coding style, readability, or security.
Models may generate functionally correct but suboptimal or unsafe code, which is not penalized under the current evaluation framework.

Despite these limitations, we believe DSCodeBench offers a strong and necessary step toward realistic, large-scale, and rigorous evaluation of LLMs for real-world data science programming and provides a valuable foundation for further benchmark development.

\subsection{Improvement Strategy}
\label{appendix: Improvement Strategy}

To address the current limitations of DSCodeBench and further enhance its utility for the community, we outline several concrete improvement strategies for future iterations of the benchmark.

\paragraph{Expanding Language and Library Coverage.}
We plan to broaden DSCodeBench beyond Python by incorporating tasks from additional programming languages commonly used in data science and scientific computing, such as R.
Moreover, we aim to include tasks that leverage emerging libraries and frameworks to better capture evolving trends in the data science ecosystem.

\paragraph{Restoring Realistic Error Handling and Complex Workflows.}
To increase the realism of coding scenarios, we will revisit the current simplification of error-handling logic.
Future versions of DSCodeBench will retain authentic error-raising behaviors and incorporate exception management patterns, enabling models to demonstrate robustness in handling imperfect inputs and edge cases. 
Additionally, we intend to expand beyond single-file tasks by introducing multi-function and multi-module coding problems, bridging the gap toward project-level code generation.

\paragraph{Broadening Evaluation Dimensions.}
Although DSCodeBench currently focuses on functional correctness, future evaluations will also consider additional quality metrics, such as runtime efficiency, code readability, adherence to best practices, and basic security properties.
Incorporating these dimensions will enable a more holistic evaluation of LLM-generated code and better reflect real-world coding standards.

Through these improvement strategies, we aim to make DSCodeBench not only more comprehensive and realistic but also more reflective of the diverse competencies required in real-world data science programming.

\subsection{Broader Impact}
\label{appendix: Broader Impact}

DSCodeBench is designed to advance the evaluation of LLMs for real-world data science code generation, and its broader impacts span research, education, and industry.

\paragraph{Promoting Realistic and Rigorous Evaluation.}
By providing a benchmark rooted in real-world coding scenarios with rigorous testing and diverse problem contexts, DSCodeBench encourages the development of LLMs that move beyond solving synthetic or overly simplified problems.
This shift helps align research progress with practical requirements faced by data scientists, researchers, and engineers in professional settings.

\paragraph{Improving LLM Robustness and Practicality.}
Through its complexity and breadth, DSCodeBench reveals areas where current LLMs struggle, such as handling complex data workflows, generating robust and generalizable solutions, and reasoning over longer contexts.
As models improve on DSCodeBench, they are more likely to develop skills needed for reliable deployment in real-world data science and engineering workflows, ultimately leading to safer and more productive AI-assisted coding tools.

\paragraph{Supporting Education and Training.}
DSCodeBench can also serve as a resource for education and training, helping instructors and students explore the capabilities and limitations of LLMs in data science contexts.
Benchmark examples drawn from realistic workflows offer valuable case studies for teaching practical coding skills, evaluating model behavior, and fostering critical discussions about AI in programming.

\paragraph{Potential Risks and Ethical Considerations.}
Despite its benefits, widespread reliance on benchmarks like DSCodeBench could introduce risks if models are over-optimized for benchmark performance without improving underlying reasoning or robustness.
Furthermore, increased automation in code generation, while enhancing productivity, may lead to skill atrophy among developers if used indiscriminately.
To mitigate these risks, we emphasize that DSCodeBench should be used alongside diverse evaluation methods and in settings that promote responsible, human-centered deployment of AI tools.

In summary, we believe DSCodeBench will have a positive impact on the field by raising the standards for evaluating LLMs in real-world programming tasks while simultaneously opening important discussions around responsible and human-centered AI development.

\subsection{Randomness Control}
\label{appendix: Randomness Control}

\paragraph{Random Seed} 
In our evaluation framework, we set the random seed to 42 to maintain the consistency of identical experiments.
For some libraries, such as NumPy, Tensorflow, and Pytorch, we can easily set the random seed by using their default API function, like \textit{np.random.seed()}, \textit{tf.random.set\_seed()}, and \textit{torch.manual\_seed()}.
But there are also libraries, such as Sklearn, which do not have specific API functions to control the global random seed.
Instead, they control randomness by setting the value of \textit{random\_state} in the API functions, such as \textit{RandomForestClassifier(random\_state=None)}.
For these situations, we explicitly reset the value of \textit{random\_state} to 42 by iterating through the AST nodes of the code.

\paragraph{Temperature} 
In our experiment, we set the temperature to 0.2.
In addition, we run the experiments on the open-source model with temperature=0.6.

Table~\ref{table: DSCodeBench different temperature} demonstrates how temperature impacts various LLMs on DSCodeBench.
For DeepSeek models, increasing temperature from 0.2 to 0.6 often results in a slight decrease in pass@3 (e.g., from 0.195 to 0.214 for the DeepSeek-Coder-6.7B-Instruct) and a modest drop or stability in the average number of correct code generated, suggesting a controlled diversity gain.
In contrast, Qwen2.5 models show a more volatile response.
For example, the Qwen2.5-Coder-7B-Instruct and Qwen2.5-Coder-14B-Instruct models see a significant drop in pass@1 and correctness at temperature 0.6, indicating a degradation in performance due to over-randomization.
However, the Qwen2.5-Coder-32B-Instruct model maintains high pass@3 performance at both temperatures, showing robustness.
These nuanced shifts demonstrate that DSCodeBench can still follow the scaling law, where larger models generally perform better than smaller ones, even under varying temperature settings, highlighting its robustness and effectiveness in evaluating LLM performance.

Table~\ref{table: DSCodeBench experiment each library correct 0.6 (mean+std)} and Table~\ref{table: DSCodeBench experiment each library 0.6 (pass@k)} showcase the per-library performance of various LLMs on DSCodeBench at temperature 0.6, highlighting both the number of fully correct code generations and pass@1 across ten major data science libraries.
DeepSeek models, especially the V2-Lite and 33B variants, exhibit strong results across libraries like NumPy, Scikit-learn, TensorFlow, and PyTorch, while Qwen2.5 models show more variability, with the 32B model partially closing the performance gap.
The results also reveal library-specific difficulty—LightGBM, Seaborn, and Matplotlib consistently yield lower performance across all models, indicating either higher complexity or lower representation in model training.
These findings underscore DSCodeBench’s strength as a fine-grained diagnostic benchmark capable of capturing nuanced model behaviors across subdomains.
It not only reflects overall model performance but also exposes domain-specific weaknesses, making it a powerful tool for evaluating and advancing LLMs in real-world data science code generation.

\begin{table*}[ht]\small
\caption{Experiment result on DSCodeBench for different temperatures.}
\vspace{0mm}
\centering
\resizebox{\linewidth}{!}{
\begin{tabular}{l l r r r r r}
\toprule
Temperature & Model & pass@1 & pass@3 &Avg. count\_correct & Avg. count\_part\_correct & Avg. count\_wrong \\
\midrule
\multirow{7}{*}{0.2} & DeepSeek-Coder-1.3B-Instruct & 0.076 & 0.103 & 76.3$\pm$3.3 & 27.0$\pm$2.2 & 896.7$\pm$5.4 \\
& DeepSeek-Coder-6.7B-Instruct & 0.163 & 0.195 & 162.7$\pm$2.1 & 47.0$\pm$1.6 & 790.3$\pm$2.1 \\
& DeepSeek-Coder-V2-Lite-Instruct(15.7B) & 0.205 & 0.234 & 205.0$\pm$1.4 & 48.0$\pm$0.8 & 747.0$\pm$0.8 \\
& DeepSeek-Coder-33B-Instruct & 0.222 & 0.258 & 222.3$\pm$2.6 & 51.0$\pm$5.4 & 726.7$\pm$4.0 \\
\cmidrule{2-7}
& Qwen2.5-Coder-7B-Instruct & 0.116 & 0.164 & 116.3$\pm$7.1 & 28.7$\pm$0.5 & 855.0$\pm$7.5 \\
& Qwen2.5-Coder-14B-Instruct & 0.213 & 0.251 & 212.7$\pm$5.6 & 49.0$\pm$1.4 & 738.3$\pm$5.3 \\
& Qwen2.5-Coder-32B-Instruct & 0.229 & 0.260 & 228.7$\pm$2.6 & 45.7$\pm$4.2 & 725.7$\pm$2.5 \\
\midrule
\midrule
\multirow{7}{*}{0.6} & DeepSeek-Coder-1.3B-Instruct & 0.057 & 0.103 & 57.3$\pm$5.2 & 24.3$\pm$7.7 & 918.3$\pm$11.8 \\
& DeepSeek-Coder-6.7B-Instruct & 0.157 & 0.214 & 157.3$\pm$11.1 & 49.7$\pm$1.2 & 793.0$\pm$9.9 \\
& DeepSeek-Coder-V2-Lite-Instruct(15.7B) & 0.196 & 0.245 & 196.3$\pm$6.8 & 54.7$\pm$4.6 & 749.0$\pm$11.0 \\
& DeepSeek-Coder-33B-Instruct & 0.203 & 0.265 & 202.7$\pm$6.8 & 46.0$\pm$3.7 & 746.3$\pm$5.6 \\
\cmidrule{2-7}
& Qwen2.5-Coder-7B-Instruct & 0.028 & 0.054 & 28.3$\pm$2.5 & 12.0$\pm$1.6 & 959.7$\pm$4.1 \\
& Qwen2.5-Coder-14B-Instruct & 0.055 & 0.081 & 54.7$\pm$2.4 & 25.3$\pm$2.6 & 920.0$\pm$2.8 \\
& Qwen2.5-Coder-32B-Instruct & 0.159 & 0.267 & 158.7$\pm$65.2 & 41.7$\pm$6.6 & 799.7$\pm$71.1 \\

\bottomrule
\end{tabular}
}
\label{table: DSCodeBench different temperature}
\end{table*}

\begin{table*}[ht]
\caption{Experiment result on DSCodeBench for each library at temperature=0.6 (number of code solutions correctly passing
all the test cases, the results are shown in the format of mean±standard deviation).}
\vspace{0mm}
\centering
\resizebox{\linewidth}{!}{
\begin{tabular}{l r r r r r r r r r r}
\toprule
Model & NumPy & Pandas & Scipy & Scikit-learn & Matplotlib & Seaborn & TensorFlow & PyTorch & Keras & LightGBM \\
\midrule
DeepSeek-Coder-1.3B-Instruct & 9.3$\pm$0.5 & 2.7$\pm$0.5 & 1.3$\pm$0.5 & 14.7$\pm$2.1 & 2.0$\pm$0.8 & 3.3$\pm$0.9 & 6.7$\pm$4.0 & 12.7$\pm$3.3 & 2.3$\pm$1.7 & 2.3$\pm$0.5\\
DeepSeek-Coder-6.7B-Instruct & 25.3$\pm$3.1 & 7.0$\pm$0.0 & 3.7$\pm$0.5 & 34.3$\pm$1.2 & 3.0$\pm$0.0 & 8.0$\pm$0.0 & 32.7$\pm$5.2 & 29.7$\pm$4.1 & 11.0$\pm$0.0 & 2.7$\pm$0.5\\
DeepSeek-Coder-V2-Lite-Instruct(15.7B) & 25.7$\pm$0.9 & 13.0$\pm$0.8 & 3.0$\pm$0.0 & 42.3$\pm$1.2 & 2.3$\pm$0.5 & 8.7$\pm$0.5 & 36.3$\pm$2.9 & 42.3$\pm$3.3 & 16.7$\pm$1.2 & 6.0$\pm$0.0\\
DeepSeek-Coder-33B-Instruct & 32.3$\pm$5.2 & 10.0$\pm$1.6 & 3.0$\pm$0.8 & 41.7$\pm$1.7 & 2.3$\pm$0.5 & 7.7$\pm$0.5 & 41.0$\pm$2.8 & 37.0$\pm$5.0 & 12.3$\pm$0.5 & 5.3$\pm$0.5\\
\midrule
Qwen2.5-Coder-7B-Instruct & 18.3$\pm$3.1 & 6.0$\pm$1.4 & 1.0$\pm$0.8 & 0.0$\pm$0.0 & 0.0$\pm$0.0 & 0.0$\pm$0.0 & 0.0$\pm$0.0 & 0.0$\pm$0.0 & 0.0$\pm$0.0 & 3.0$\pm$0.0\\
Qwen2.5-Coder-14B-Instruct & 30.7$\pm$2.1 & 17.7$\pm$1.2 & 2.3$\pm$0.5 & 0.0$\pm$0.0 & 1.0$\pm$0.0 & 0.0$\pm$0.0 & 0.0$\pm$0.0 & 0.0$\pm$0.0 & 0.0$\pm$0.0 & 3.0$\pm$0.0\\
Qwen2.5-Coder-32B-Instruct & 37.0$\pm$0.8 & 16.3$\pm$2.6 & 3.0$\pm$0.0 & 20.0$\pm$14.2 & 1.0$\pm$0.0 & 3.7$\pm$2.6 & 28.3$\pm$20.2 & 29.7$\pm$21.0 & 12.0$\pm$8.6 & 7.7$\pm$1.9\\

\bottomrule
\end{tabular}
}
\label{table: DSCodeBench experiment each library correct 0.6 (mean+std)}
\end{table*}

\begin{table*}[ht]\tiny
\caption{Experiment result on DSCodeBench for each library at temperature=0.6 (pass@1).}
\vspace{0mm}
\centering
\resizebox{\linewidth}{!}{
\begin{tabular}{l r r r r r r r r r r}
\toprule
Model & NumPy & Pandas & Scipy & Scikit-learn & Matplotlib & Seaborn & TensorFlow & PyTorch & Keras & LightGBM \\
\midrule
DeepSeek-Coder-1.3B-Instruct & 0.071 & 0.029 & 0.012 & 0.136 & 0.019 & 0.040 & 0.061 & 0.125 & 0.022 & 0.043\\
DeepSeek-Coder-6.7B-Instruct & 0.193 & 0.076 & 0.033 & 0.318 & 0.029 & 0.096 & 0.297 & 0.294 & 0.106 & 0.049\\
DeepSeek-Coder-V2-Lite-Instruct(15.7B) & 0.196 & 0.141 & 0.027 & 0.392 & 0.022 & 0.104 & 0.330 & 0.419 & 0.160 & 0.111\\
DeepSeek-Coder-33B-Instruct & 0.247 & 0.109 & 0.027 & 0.386 & 0.022 & 0.092 & 0.373 & 0.366 & 0.119 & 0.099\\
\midrule
Qwen2.5-Coder-7B-Instruct & 0.140 & 0.065 & 0.009 & 0.000 & 0.000 & 0.000 & 0.000 & 0.000 & 0.000 & 0.056\\
Qwen2.5-Coder-14B-Instruct & 0.234 & 0.192 & 0.021 & 0.000 & 0.010 & 0.000 & 0.000 & 0.000 & 0.000 & 0.056\\
Qwen2.5-Coder-32B-Instruct & 0.282 & 0.178 & 0.027 & 0.185 & 0.010 & 0.044 & 0.258 & 0.294 & 0.115 & 0.142\\
\bottomrule
\end{tabular}
}
\label{table: DSCodeBench experiment each library 0.6 (pass@k)}
\end{table*}

\subsection{Alignment}
\label{appendix: Alignment}

To ensure the reliability of our benchmark, alignment between the three core components—test cases, ground truth code, and code problem descriptions—is critical. 

First, we ensure alignment between test cases and ground truth code by using test case generation scripts tailored to the ground truth implementation. 
All generated test cases must pass the ground truth code; otherwise, we iteratively prompt LLMs to refine the generation script.
We further validate sufficiency by conducting test case coverage analysis, quantifying how thoroughly the test cases can cover the code logic branches (details in Appendix: Test Case Coverage). 

Second, to align test cases with code problem descriptions, we manually ensure that the input/output formats in the problem description precisely match the parameter structure used in the test case generation script. 
Additionally, example input-output pairs are directly generated randomly from the test case generation scripts. 

Third, we align the ground truth code and problem description by generating the latter via LLMs and filtering out misaligned cases during the pipeline.
In the final manual editing, we define alignment based on whether a human developer could correctly implement the ground truth solution based on the problem description, ensuring the benchmark's clarity and usability.
We employ both human experts (two of our authors) and LLMs (GPT-4o-mini and GPT-4o) as judges to assess alignment, achieving 97.4\% agreement among all the tasks.
For the remaining 2.6\%, two authors conducted targeted manual corrections.

\subsection{Data Leakage Mitigation}
\label{appendix: Data Leakage Mitigation}

Although DSCodeBench is constructed from source code retrieved from GitHub, which inherently introduces potential data leakage risks, we have taken multiple measures to mitigate this issue throughout the construction pipeline.
Specifically, we avoid directly reusing the original code snippets by reconstructing the code context and performing careful manual editing to ensure the resulting problems differ substantially from the source. 

\begin{figure}
\centerline{\includegraphics[width=0.8\linewidth]{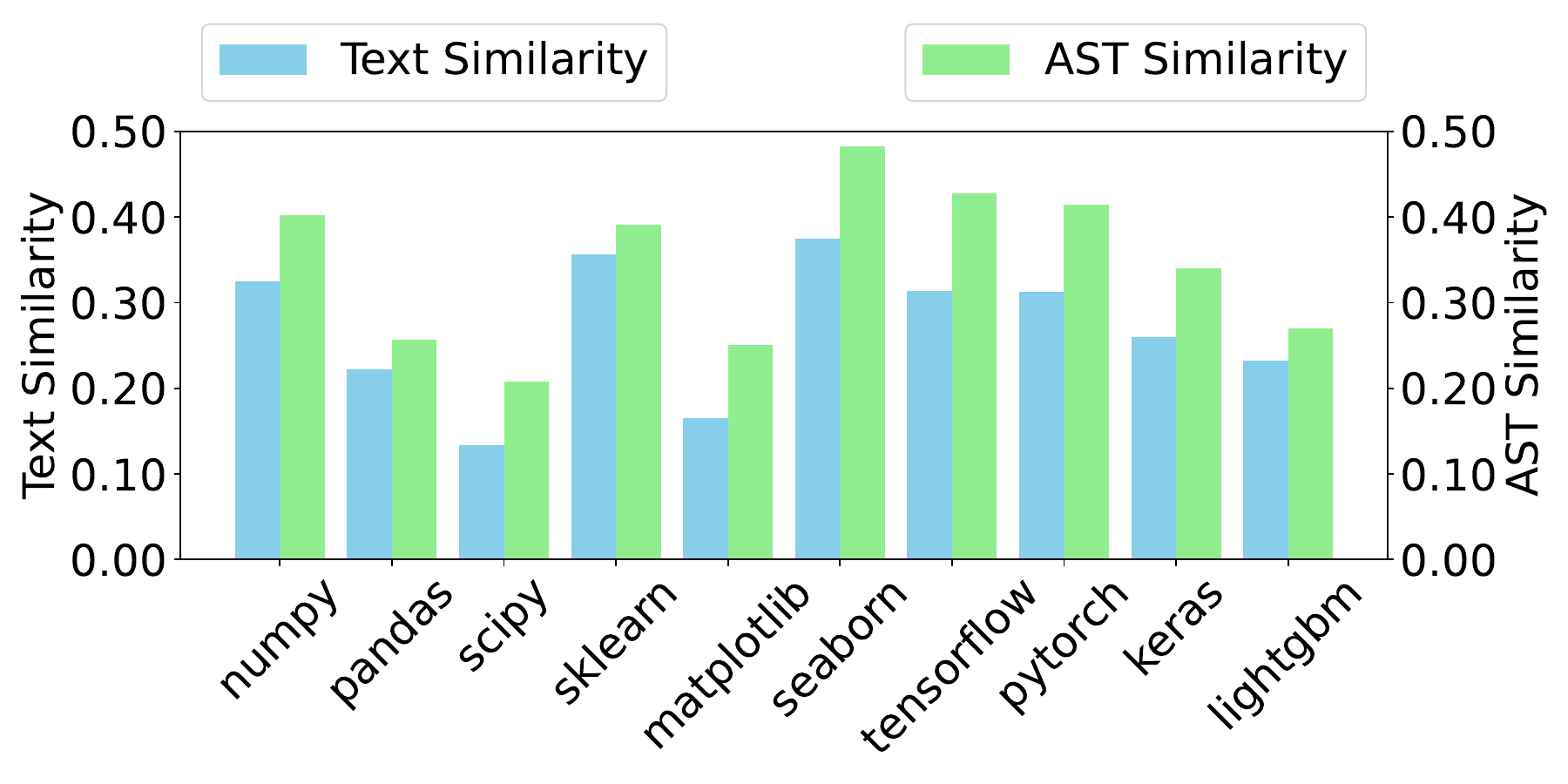}}
\vspace{0mm}
\caption{Similarity between LLM-generated solution and ground truth code by different libraries.}
\label{fig: Similarity_library}
\end{figure}

\begin{figure}
\centerline{\includegraphics[width=0.8\linewidth]{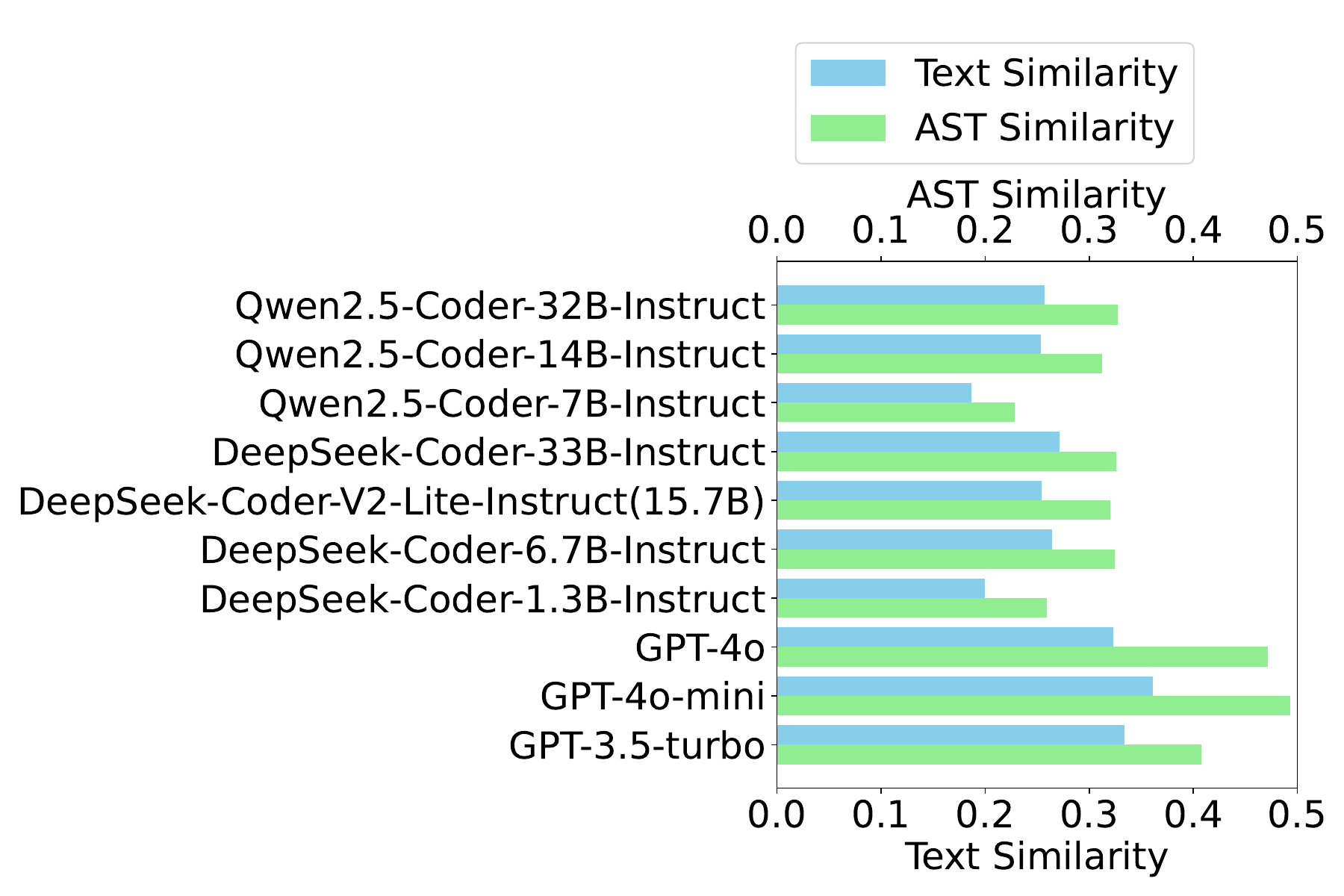}}
\vspace{0mm}
\caption{Similarity between LLM-generated solution and ground truth code by different models.}
\label{fig: Similarity_model}
\end{figure}

To assess the severity of any remaining leakage, we conduct a comprehensive similarity analysis between the LLM-generated code and the ground truth code across three dimensions: semantic similarity, syntactic similarity, and structural similarity. 
Semantically, the pass@k performance of LLMs is significantly lower than the ideal 100\% pass@k of ground truth code, indicating a lack of direct memorization.
Syntactically, we compute text-level similarity and observe scores consistently below 0.4 across different libraries and models.
Structurally, we evaluate AST-based similarity and find values below 0.5, further confirming minimal overlap. 
Figure~\ref{fig: Similarity_library} and Figure~\ref{fig: Similarity_model} show the similarity distributions between LLM-generated code and ground truth code across different libraries and different models, respectively.
These results collectively demonstrate that DSCodeBench has made strong efforts to mitigate data leakage and preserve benchmark integrity.

\subsection{Test Case Coverage}
\label{appendix: Test Case Coverage}

To demonstrate the quality and completeness of our automatically generated test case scripts, we conducted a comprehensive test case coverage experiment. 
The purpose of this experiment is to assess how well our test case scripts can cover the logic of the ground truth code, which ultimately serves to test the correctness of LLM-generated code.
The experiment was conducted as follows.

For each problem, the ground truth code and the corresponding test case script were saved into two separate files.
We developed an auxiliary script that first executes the main function in the test case script to automatically generate 200 test cases.
These test cases were then systematically fed into the main function of the ground truth code, simulating the full verification workflow of LLM-generated code.
To measure how thoroughly the test cases exercised the code paths of the ground truth implementation, we employed the widely used Python library \textit{coverage}.
The coverage tool provides a detailed report of code execution, from which we recorded the final line coverage percentage for each problem.

\begin{figure}
\centerline{\includegraphics[width=0.8\linewidth]{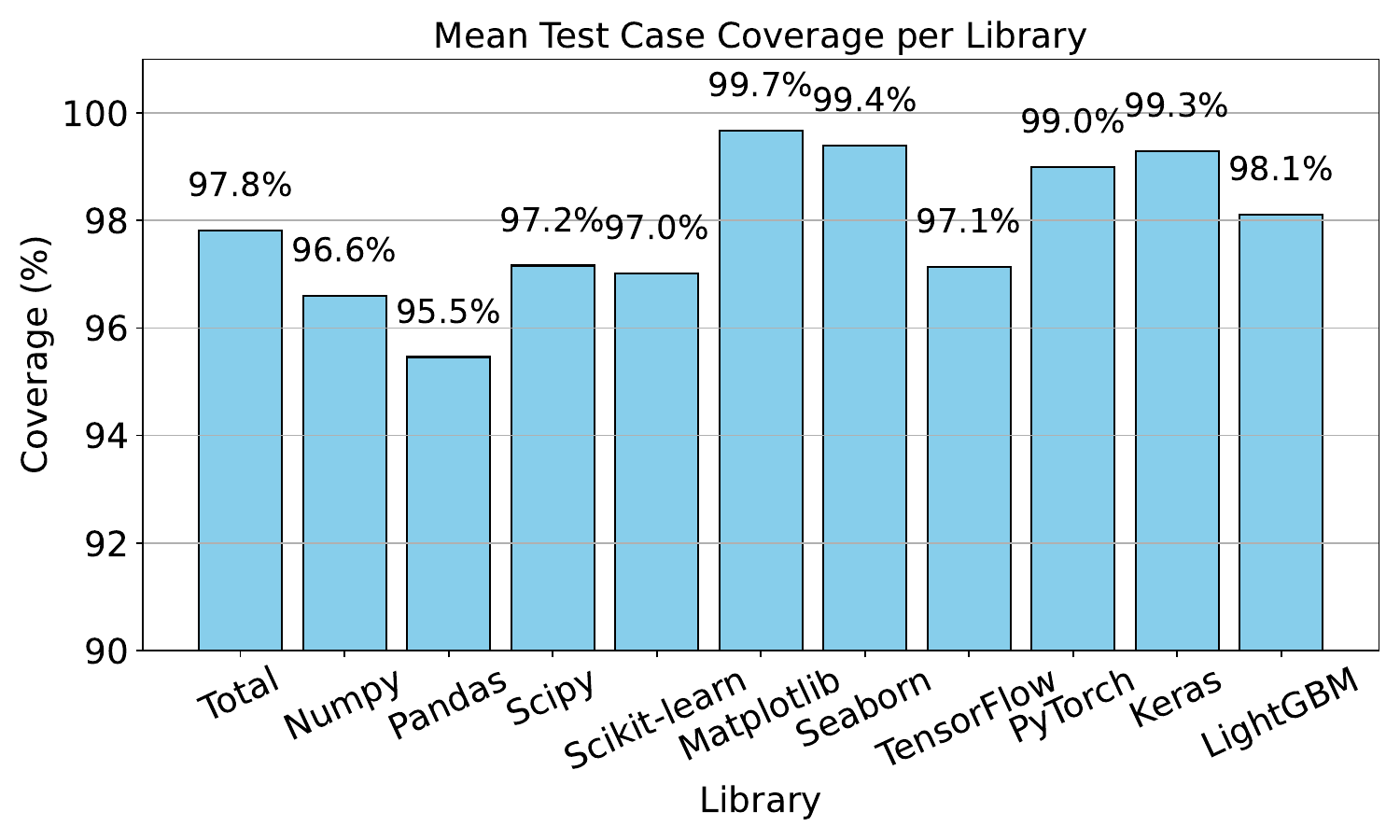}}
\vspace{0mm}
\caption{Mean test case coverage on each library.}
\label{fig: coverage}
\end{figure}

We aggregated the results and calculated the mean coverage for each library as well as the overall mean across all problems. 
Figure~\ref{fig: coverage} presents the test case coverage results. 
The overall mean coverage across all problems reaches 97.8\%, indicating that our test case scripts are highly effective in exercising the ground truth code.

Analyzing individual libraries, we observe that the scientific plotting libraries, Matplotlib (99.7\%) and Seaborn (99.4\%), achieved the highest coverage.
Deep learning frameworks PyTorch (99.3\%), TensorFlow (99.0\%), and Keras (99.3\%) also show excellent coverage, reflecting the strong compatibility between our test case generation script and ground truth code. 
Meanwhile, data manipulation libraries such as Numpy (96.6\%), Pandas (95.5\%), and Scipy (97.2\%) report slightly lower coverage.
Overall, the consistently high coverage across all libraries demonstrates the robustness and reliability of our test case generation pipeline and its suitability for evaluating LLM-generated code in realistic, diverse data science scenarios.

\subsection{Test Suite Design}
\label{appendix: Test Suite Design}

In our benchmark, the test suite is systematically divided into two categories: plot-drawing tasks and non-plotting tasks.
For plot-drawing libraries, including Matplotlib and Seaborn, we evaluate the similarity between the generated and reference images using the structural\_similarity function from skimage.metrics. 
A test case is considered successful if the mean similarity score across the RGB channels exceeds 0.5; otherwise, it is deemed to have failed. 
For non-plotting libraries, we first identify the return type of the output. 
Common return types include standard Python types such as lists, dictionaries, and numeric types (e.g., integers or floats), as well as library-specific types such as tf.Tensor, pd.DataFrame, and torch.nn.Sequential. 
When the return values are directly comparable, we perform a direct value comparison. 
In cases where direct comparison is not feasible, we extract and compare relevant attributes of the returned objects, as exemplified by the attribute-wise comparison of torch.nn.Sequential instances.

\subsection{Comparison between DS-1000 and DSCodeBench}
\label{appendix: Comparison between DS-1000 and DSCodeBench}

DS-1000 and DSCodeBench differ significantly in terms of task complexity, clarity, and robustness of evaluation.
DS-1000 tasks are loosely defined, often derived from informal user queries (e.g., StackOverflow), leading to ambiguous problem statements, inconsistent formats, and limited input diversity.

The following examples compare the code problem description, code solutions, and test case scripts between DS-1000 and DSCodeBench.
The DS-1000 code solution is short and focuses on a single function call (\textit{CrossEntropyLoss}), with almost no surrounding logic or structure.
In contrast, the DSCodeBench code solution presents a more complex scenario, requiring more steps, including vector normalization, matrix operations (\textit{einsum}), and parameter scaling.
This reflects DSCodeBench’s emphasis on more realistic and multi-step coding tasks that better mimic real-world development workflows.
The test case scripts further highlight the differences.
DS-1000 provides only a single hard-coded test case, with fixed random seeds and dimensions, offering minimal variability and limited robustness for model evaluation.
By comparison, DSCodeBench introduces a test case generator function that, by default, generates 200 diverse input samples, with test cases spanning a much wider range of values and supporting customizable random seeds.
This test case design ensures better coverage of potential input spaces, promoting a more robust model evaluation.
The code problem descriptions in DS-1000 often suffer from excessive verbosity, informal tone, and ambiguous requirements, which can lead to inconsistent model behavior. 
Many prompts are long, poorly formatted, and derived from conversational or irrelevant content in online forums, resulting in vague task objectives and unclear constraints.
This reflects the unstructured nature of DS-1000 and poses significant challenges for reliable evaluation. 
In contrast, DSCodeBench provides clear, well-structured prompts aligned with real-world data science workflows. 
Each problem includes a precise function signature, a detailed breakdown of the implementation logic, and explicit input-output specifications, enabling deterministic evaluation and improving reproducibility.



\begin{DS1000Box}[title=\textbf{Code Problem Description (DS1000)}]
Problem:
I am doing an image segmentation task. There are 7 classes in total so the final outout is a tensor like [batch, 7, height, width] which is a softmax output. Now intuitively I wanted to use CrossEntropy loss but the pytorch implementation doesn't work on channel wise one-hot encoded vector
So I was planning to make a function on my own. With a help from some stackoverflow, My code so far looks like this
\begin{lstlisting}[language=Python, numbers=none]
from torch.autograd import Variable
import torch
import torch.nn.functional as F
def cross_entropy2d(input, target, weight=None, size_average=True):
    # input: (n, c, w, z), target: (n, w, z)
    n, c, w, z = input.size()
    # log_p: (n, c, w, z)
    log_p = F.log_softmax(input, dim=1)
    # log_p: (n*w*z, c)
    log_p = log_p.permute(0, 3, 2, 1).contiguous().view(-1, c)  # make class dimension last dimension
    log_p = log_p[
       target.view(n, w, z, 1).repeat(0, 0, 0, c) >= 0]  # this looks wrong -> Should rather be a one-hot vector
    log_p = log_p.view(-1, c)
    # target: (n*w*z,)
    mask = target >= 0
    target = target[mask]
    loss = F.nll_loss(log_p, target.view(-1), weight=weight, size_average=False)
    if size_average:
        loss /= mask.data.sum()
    return loss
images = Variable(torch.randn(5, 3, 4, 4))
labels = Variable(torch.LongTensor(5, 4, 4).random_(3))
cross_entropy2d(images, labels)
\end{lstlisting}

\vspace{\baselineskip}
I get two errors. One is mentioned on the code itself, where it expects one-hot vector. The 2nd one says the following
RuntimeError: invalid argument 2: size `[5 x 4 x 4 x 1]' is invalid.
For example purpose I was trying to make it work on a 3 class problem. So the targets and labels are (excluding the batch parameter for simplification ! )\\
(test case is omitted here for brevity)\\
So how can I fix my code to calculate channel wise CrossEntropy loss ?
Or can you give some simple methods to calculate the loss? Thanks
Just use the default arguments\\
A:\\
\begin{lstlisting}[language=Python, numbers=none]
<code>
import numpy as np
import pandas as pd
from torch.autograd import Variable
import torch
import torch.nn.functional as F
images, labels = load_data()
</code>
loss = ... # put solution in this variable
BEGIN SOLUTION
<code>
\end{lstlisting}

\end{DS1000Box}

\begin{DSCodeBenchBox}[title=\textbf{Code Problem Description (DSCodeBench)}]
You are tasked with implementing a function that computes the contrastive loss between two sets of embeddings. The function should normalize the input embeddings, compute the logits using the dot product of the normalized embeddings, and then apply the cross-entropy loss to obtain the final loss value. The function signature is as follows: \\

\begin{lstlisting}[language=Python, numbers=none]
def contrastive_loss(q: torch.Tensor, k: torch.Tensor, T: float) -> torch.Tensor:
\end{lstlisting}

\vspace{\baselineskip}

In this function:\\
- `q` is a tensor representing the first set of embeddings.
- `k` is a tensor representing the second set of embeddings.
- `T` is a float constant that acts as a temperature parameter for scaling the logits.
The constant used in the main code is `2 * T`, which is used to scale the final loss value.\\

The logic of the main code is as follows:\\

1. Normalize the input tensors `q` and `k` along the specified dimension (dim=1). This ensures that each embedding has a unit norm, which is essential for contrastive learning.

2. Compute the logits by calculating the dot product of the normalized embeddings using the Einstein summation convention (`torch.einsum`). The logits are scaled by dividing by the temperature parameter `T`.

3. Determine the number of samples `N` from the shape of the logits tensor.

4. Create a tensor `labels` that contains the indices of the samples, which will be used as the target labels for the cross-entropy loss. This tensor is created using `torch.arange(N)`.

5. Compute the cross-entropy loss using `nn.CrossEntropyLoss()`, passing in the logits and the labels. This loss measures how well the model's predictions (logits) match the true labels.

6. Finally, multiply the computed loss by `2 * T` to scale the loss appropriately before returning it.\\

Input format:\\
- `q`: A tensor of shape (N, D) where N is the number of samples and D is the dimensionality of the embeddings.
- `k`: A tensor of shape (N, D) where N is the number of samples and D is the dimensionality of the embeddings.
- `T`: A float representing the temperature parameter.\\

Output format:\\
- Returns a tensor representing the scaled contrastive loss value. \\

Input: \\
\begin{lstlisting}[language=Python, numbers=none]
q = torch.tensor([[ 0.5, -0.2,  0.1],
                  [ 0.3,  0.4, -0.6],
                  [-0.1,  0.8,  0.2]])
k = torch.tensor([[ 0.4, -0.1,  0.3],
                  [ 0.2,  0.5, -0.7],
                  [-0.3,  0.6,  0.1]])
T = 0.5
\end{lstlisting}

\vspace{\baselineskip}
Output:
\begin{lstlisting}[language=Python, numbers=none]
loss = contrastive_loss(q, k, T)
# loss = tensor(0.2475)
\end{lstlisting}

\vspace{\baselineskip}

\end{DSCodeBenchBox}

\begin{DS1000Box}[title=\textbf{Ground Truth Code (DS1000)}]
\begin{lstlisting}[language=Python, numbers=none]
def generate_ans(data):
    images, labels = data
    loss_func = torch.nn.CrossEntropyLoss()
    loss = loss_func(images, labels)
    return loss
\end{lstlisting}

\end{DS1000Box}

\begin{DSCodeBenchBox}[title=\textbf{Ground Truth Code (DSCodeBench)}]
\begin{lstlisting}[language=Python, numbers=none]
def contrastive_loss(q, k, T):
    q = nn.functional.normalize(q, dim=1)
    k = nn.functional.normalize(k, dim=1)
    logits = torch.einsum('nc,mc->nm', [q, k]) / T
    N = logits.shape[0]
    labels = torch.arange(N, dtype=torch.long)
    return nn.CrossEntropyLoss()(logits, labels) * (2 * T)
\end{lstlisting}
\end{DSCodeBenchBox}

\begin{DS1000Box}[title=\textbf{Test Case Script (DS1000)}]
\begin{lstlisting}[language=Python, numbers=none]
def define_test_input(test_case_id):
    if test_case_id == 1:
        torch.random.manual_seed(42)
        images = torch.randn(5, 3, 4, 4)
        labels = torch.LongTensor(5, 4, 4).random_(3)
    return images, labels
\end{lstlisting}
\end{DS1000Box}

\begin{DSCodeBenchBox}[title=\textbf{Test Case Script (DSCodeBench)}]
\begin{lstlisting}[language=Python, numbers=none]
def test_case_input_generator(n=200):
    test_cases = []
    for _ in range(n):
        batch_size = random.randint(2, 128)
        feature_dim = random.randint(16, 512)
        T = random.uniform(0.01, 1.0)
        q = torch.randn(batch_size, feature_dim)
        k = torch.randn(batch_size, feature_dim)
        test_cases.append((q, k, T))
    return test_cases
\end{lstlisting}
\end{DSCodeBenchBox}

\subsection{Prompt Format}
\label{appendix: Prompt Format}

Several stages of our benchmark construction process involve the use of LLMs.
First, we leverage LLMs to generate test case scripts.
Second, we use LLMs to generate natural language problem descriptions from ground truth code.
Finally, for alignment behavior analysis, we prompt the LLM to generate code solutions based on the given problem descriptions.
In our experiments, we also use the prompt that was used in alignment behavior analysis to generate code solutions for benchmarking the state-of-the-art LLMs.
The prompts we used in this paper are shown as follows.

\begin{PromptBox}[title=\textbf{Code Problem Description Generation Prompt}, label={box:Code Problem Description Generation Prompt}]
Please generate a code problem description of the following code. The description must include the function signature of the main code. Please specify the constant used in the main code. Please explain the logic detail step by step in the main code. Please provide input and output format of the code, but do not provide any input or output examples. Do not provide constrains. Do not provide any utility code. All the code must be able to be generated based on the description standalone.\\

\# Code \# \\
\textbf{Here is the code.} \\

\# Response \# \\
The return should follow the following format (replace \{\} into the code problem): \\
Code problem description:\\
\{\}
\end{PromptBox}

\begin{PromptBox}[title=\textbf{Code Generation Prompt}]
Please generate Python3 solution for the following code problem description: \\

\# Code problem description \# \\
\textbf{Here is the code problem description.} \\

\# Response \# \\
The return should follow the following format (replace \{\} with the solution). Do not generate additional code, such as "\_\_main\_\_" block. 

Solution:

\{\}
\end{PromptBox}

\begin{PromptBox}[title=\textbf{Test Case Script Generation Prompt}]
Please generate a script to automatically generate (n=200) test cases input for the following main code function, make sure the tensor shapes match properly (please return with the following format): \\

\begin{lstlisting}[language=Python, numbers=none]
def test_case_input_generator(n=200):
   test_cases = []
   for _ in range(n):
       XXX


       test_cases.append(XXX)
   return test_cases
\end{lstlisting}
\vspace{\baselineskip}

\# Code \# \\
\textbf{Here is the code.}
\end{PromptBox}

\begin{PromptBox}[title=\textbf{Test Case Script Repair Prompt}]
Please re-generate a python function named `test\_case\_input\_generator' to generate high-quality test cases input for the following ground truth solution. The previous version of the script has an error. The script should take n (the total number of test cases input, set default as 200) as input. Do not use any example test case in the code.  

\# Ground Truth Solution \# \\
\textbf{Here is the ground truth code.} \\

\# Previous Test Case Generation Script \# \\
\textbf{Here is the test case generation script.} \\

\# Error Info \# \\
\textbf{Here is the error information.} \\

\end{PromptBox}

\begin{PromptBox}[title=\textbf{LLM judge Prompt}]
Please check whether the `Code Problem' is sufficient enough and proper to test a data science developer's capability in finishing the task. (i.e., whether a human could generate the `Solution' based on the `code problem')\\

\# Code Problem \#

\textbf{Here is the code problem description.} \\

\# Solution \#

\textbf{Here is the code.} \\

\# Return Format \# 

Please only return True or False.

If the answer is False, please state the reason on the second line, starting with `REASON:'

If the answer is False, please also suggest the new code problem based on the `Solution'
\end{PromptBox}


\end{document}